\documentclass[aps,superscriptaddress,showpacs,nofootinbib]{revtex4-1}
\usepackage{bm}
\usepackage{graphicx}

\begin{document}

\title{Multi-state transitions and quantum oscillations of optical activity}

\author{Celia Blanco}
\email{blancodtc@cab.inta-csic.es} \affiliation{Centro de
Astrobiolog\'{\i}a (CSIC-INTA), Carretera Ajalvir Kil\'{o}metro 4,
28850 Torrej\'{o}n de Ardoz, Madrid, Spain}
\author{David Hochberg}
\email{hochbergd@cab.inta-csic.es} \affiliation{Centro de
Astrobiolog\'{\i}a (CSIC-INTA), Carretera Ajalvir Kil\'{o}metro 4,
28850 Torrej\'{o}n de Ardoz, Madrid, Spain}

\begin{abstract}
We consider the effects of multi-state transitions on the tunneling
racemization of chiral molecules. This requires going beyond simple
two-state models of enantiomers and to include transitions within a
multiple level quantum mechanical system. We derive an
\textit{effective} two-level description which accounts for
transitions from the enantiomers to an arbitrary number of excited
states as an application of the Weisskopf-Wigner approximation
scheme. Modifications to the optical activity from these additional
states are considered in general terms under the assumption of
\textit{CPT} invariance and then under T invariance. Some formal
dynamical analogies between enantiomers and the neutral K-meson
system are discussed.
\end{abstract}

\date{\today}

\maketitle

\section{\label{sec:intro} Introduction}

The effects of parity violation in the two-level approximation for
the quantum dynamics of a pair of strictly isolated molecular
enantiomers \cite{Hund} was considered some years ago by Harris and
Stodolsky (HS) \cite{HSa}. Based on this model, they pointed out the
interesting possibility that tunneling should exist leading to
oscillations in the optical activity (OA). They argued moreover that
such a system should be sensitive to extremely small energies and
could be used to observe the presence of the weak interaction via
oscillations of the optical activity about a non-zero value. Their
basic idea continues to motivate detailed proposals for measuring
the optical activity \cite{Bargueno09}, a highly nontrivial pursuit.
In this vein, it is also worth mentioning the many varied
independent experimental efforts and proposals aimed at detecting
parity violation (PV) in chiral molecules. These techniques include
vibrational-rotational \cite{Letokhov}, electronic
\cite{Quack86,Bergera,Bergerb}, Mossbauer \cite{Khriplovich}, and
nuclear magnetic resonance (NMR) spectroscopy \cite{Barra}.
Proposals for measuring the PV energy difference in crystallization
\cite{Keszthelyi} and in solubility experiments \cite{Shinitzky}
have also been considered. To date, however, no effects of parity
violation in chiral molecules have been experimentally observed. The
challenge has been taken up recently by a multi-disciplinary
consortium to employ high-resolution laser spectroscopy for a
\textit{first} observation of PV in chiral molecules \cite{Darquie}.

Returning to the HS scheme, they also recognized that obstacles to
the observation and detection of these chiral oscillations in the OA
will come from interactions with the surrounding medium such as
collision effects, thus tending to induce relaxation phenomena
\cite{HSb}. Radiative processes, fundamental interactions with the
radiation field, are also important and their inclusion necessarily
implies adopting a complex \textit{multi-level} treatment
\cite{Quacka}. Indeed, the effects of both collisional and radiative
processes can be \textit{simulated} approximately by
\textit{phenomenologically} adding complex energies
$E_k-\frac{i}{2}\Gamma_k$, to the hamiltonian spectrum, where
$\Gamma_k$ denotes a decay width \cite{Quackb}. Apart from effects
of collisions between the enantiomers and a background medium (i.e.,
gas, liquid or solid), the double well model implies a tower of
excited electronic-vibrational states for the chiral molecules
\cite{Barron1986}, and transitions between these states, induced by
an external field, is expected to modify the fundamental oscillation
period and amplitude of the OA of the ground-state enantiomers.
These considerations would seem to invalidate the simplest two-state
HS description. It is clear that the multi-state nature of real
molecules should be taken into account in order to obtain a more
accurate description \cite{Quackb}.

Motivated by the above-mentioned considerations, we probe somewhat
further in the quantum-mechanical model of HS. Therefore our
objective in this paper is to consider a multi-state approach
applied to the racemization problem and demonstrate that an
\textit{effective} two-level Hamiltonian description can be derived
from perturbation theory, thus legitimizing this description. We
derive the explicit form of the energy spectrum of the reduced
two-level system in terms of the appropriate interaction
hamiltonian. Our aim is to keep the discussion as general as
possible, thus our basic assumption is the existence of a hermitian
interaction hamiltonian responsible for inducing the transitions
from the chiral molecules to higher (e.g., electronic, vibrational,
and/or rotational) levels. Given this interaction, we then adapt
straightforwardly the Weisskopf-Wigner (WW) approximation scheme
\cite{WW}, originally used for solving the line width problem in
atomic transitions \cite{Heitler}, to the case of chiral molecules
treated as a complex multi-state quantum mechanical system. The
effect of transitions is two-fold: on the one hand, they lead to
corrections to the enantiomer mass matrix, thus lifting the
enantiomeric mass degeneracy and hence yielding corrections to the
oscillation period; and on the other, they also \textit{in
principle} allow for effects of decay channels of the enantiomers.
However, in the context of real molecules, such energy-conserving
\textit{decay channels} are most likely absent, as these would
entail the fragmentation or dissociation of the enantiomers into
smaller molecular units. This notwithstanding, the WW approach
convincingly establishes a formal analogy between the
\textit{dynamics} of the racemization of molecular enantiomers and
the interference and decay effects predicted \cite{LeeOehmeYang} and
observed \cite{TDLee,Kabir} in the neutral kaon system of elementary
particle physics. These are both examples of multiple-level quantum
systems with oscillations, although the underlying physics in each
system (chiral molecules, \textit{K}-mesons) is of course radically
different.

We give a brief overview of the simplest Harris-Stodolsky two-state
approach in Sec. \ref{sec:HS}. Motivated by the multi-level nature
of real molecules pointed out above, we approach this problem by
applying the Weisskopf-Wigner approximation scheme in Sec.
\ref{sec:multi}. This permits one to derive an effective two-level
description which takes into account transitions to an arbitrary
number of multiple excited states. The ``end" result is expressed as
a compact $2 \times 2$ mass matrix which acts in the space of the
two enantiomers. The associated eigenvalues and eigenvectors depend
on whether one assumes the putative interaction hamiltonian is
\textit{CPT} or T invariant, and these invariance principles also
effect the final form of the OA and the oscillation period. Our goal
here is to derive an effective Hamiltonian description for the
\textit{transitions} to and from multiple levels, not a study
\textit{per se} of the effects that PV may itself lead to
\cite{Bakasov,Lazzeretti,Laerdahl,
Thyssen,Bergerc,Bergerd,Schwerdtfeger,Hennum}, although these can be
included in principle as part of the overall perturbation. Dynamical
analogies between enantiomers and the neutral \textit{K}-meson
system \cite{Wigner1965,Barron1986,Barron1994}, are discussed in Sec
\ref{sec:kaon}. We comment on the novelty of the approach developed
in the Discussion, and contrast it briefly to other current
theoretical works. The WW method is summarized in the Appendix.

\section{\label{sec:HS} Harris-Stodolsky two-state model}

The Hamiltonian $H$ describing the dynamics of a pair of strictly
isolated chiral molecules $|L>,|R>$ is \cite{HSa}:
\begin{equation}
H = E_0\bm{1} +\delta \sigma_x + \epsilon \sigma_z = H_0 + \epsilon
\sigma_z,
\end{equation}
where $\delta = <L|H_0|R>$ is a parameter related to barrier height
and $\epsilon$ is the energy shift due to parity violation. For
$\epsilon = 0$, the eigenvectors  $|+> = (|L> + |R>)/\sqrt{2}$ and
$|-> = (|L> -|R>)/\sqrt{2}$ are states of definite parity: $P|+> =
|+>$ and $P|-> = -|->$. The eigenvalues are given by $E_{\pm} = E_0
\pm \delta$, and level splitting is due to tunneling alone. The
system wave function obeys $i\hbar \frac{d |\Psi>}{dt} = H |\Psi>$,
and in terms of the parity basis, the general time-dependent
solution is given by
\begin{equation}
|\Psi(t)> = a e^{-i(E_0+\delta)t/\hbar}|+> +  b
e^{-i(E_0-\delta)t/\hbar}|->
\end{equation}
where the values of $a,b$ incorporate the initial condition
$|\Psi(0)>$. Thus, if we prepare the system to be initially in the
chiral state $|L>$ , then $a=b=1/\sqrt{2}$, and at any later time $t
\geq 0$ the wavefunction is given by
\begin{equation}
|\Psi(t)> =  e^{-iE_0t/\hbar}\big(\cos(\delta t/\hbar)|L> -
i\sin(\delta t/\hbar)|R> \big).
\end{equation}
Then $P_L (t) =|<L|\Psi(t)>|^2 = \cos^2(\frac{\delta t}{\hbar})$ and
$P_R (t)= |<R|\Psi(t)>|^2= \sin^2(\frac{\delta t}{\hbar})$ are the
probabilities for the system to be in chiral state L or to make a
transition to the state R, at time $t$, respectively. Note that $P_L
(t)+ P_R (t) = 1$. The optical activity is given by \cite{HSa}
\begin{eqnarray}\label{HSOA1}
\Theta(t) &=& \Theta_{max}(P_L(t)-P_{R}(t))\nonumber \\
&=& \Theta_{max}\cos\big(\frac{2 \delta t}{\hbar}\big).
\end{eqnarray}
For parity violation, $|\epsilon| > 0$, and the eigenvalues are
$E_{1,2} = E_0 \pm (\delta^2 + \epsilon^2)^{1/2}$ and the associated
eigenstates are given by $|\Psi_1> = \cos \phi |L> + \sin \phi |R>$
and  $|\Psi_2> = -\sin \phi |L> + \cos \phi |R>$, with mixing angle
defined via $\cot 2\phi = \frac{\epsilon}{\delta}$ \cite{HSa}. If
the system is initially prepared to be in the chiral state $|L>$,
then at any later time its wavefunction is given by
\begin{equation}
|\Psi(t)> = e^{-iE_0 t/\hbar}\Big( (\cos^2\phi \, e^{-i \Delta
t/\hbar} + \sin^2 \phi \, e^{+i \Delta t/\hbar})|L> + \cos \phi \sin
\phi (e^{-i \Delta t/\hbar} - e^{i \Delta t/\hbar})|R> \Big),
\end{equation}
where $\Delta \equiv (\delta^2 + \epsilon^2)^{1/2}$. The probability
to make a transition to a state $|R>$, is therefore given by
\begin{eqnarray}\label{LtoRtrans}
P_R(t) &=& |<R|\Psi(t)>|^2 \nonumber \\
&=& 4 \sin^2 \phi \cos^2 \phi \, \sin^2\big( \sqrt{\epsilon^2 +
\delta^2} t/\hbar\big)\nonumber \\
&=& \frac{\delta^2}{\epsilon^2 + \delta^2} \sin^2\big(
\sqrt{\epsilon^2 + \delta^2} t/\hbar\big).
\end{eqnarray}
The probability to be in the state $|L>$ is:
\begin{eqnarray}
P_{L}(t) &=& |<L|\Psi(t)>|^2 \nonumber \\
&=& \cos^2(\frac{\Delta t}{\hbar}) + \cos^2 2 \phi
\,\sin^2(\frac{\Delta t}{\hbar}),\nonumber \\
&=& \cos^2(\frac{\Delta t}{\hbar}) + \frac{\epsilon^2}{\delta^2 +
\epsilon^2}\,\sin^2(\frac{\Delta t}{\hbar}).
\end{eqnarray}
Once again, we have $P_L (t)+ P_R(t) = 1$. The corresponding optical
activity (OA) is then calculated to be
\begin{eqnarray}\label{HSOA}
\Theta(t) &=& \Theta_{max}(P_L(t)-P_{R}(t))\nonumber \\
&=& \Theta_{max}\frac{\epsilon^2 + \delta^2 \cos(2 \Delta
t/\hbar)}{\delta^2 + \epsilon^2}.
\end{eqnarray}
This latter formula is the starting point for recent proposals for
measuring the parity violating energy difference $\epsilon$ between
enantiomers \cite{Bargueno09}.

\section{\label{sec:multi} Multi-level problem: transitions to many states}

We start with mass degenerate pair $|L>,|R>$ of chiral enantiomers:
$m_L = m_R \equiv m$, considered as ground states of a Hamiltonian
$H_0$. We then include transitions from these states to other
eigenstates or levels of $H_0$ induced by external fields
(radiation, thermal effects, etc.). The problem we thus consider is
the time evolution of a state initially prepared as a superposition
of $|L>$ and $|R>$. We regard the potential well barrier, the
parity-violating energy difference $\epsilon$ as part of the overall
perturbation $H_1$. So we decompose $H_1 = \delta \sigma_x +
\epsilon \sigma_z + H_2$, where the first two terms act only within
the two-dimensional subspace of the ground-state enantiomers, and
$H_2$ induces transitions from these to the other
(electronic-vibrational) levels. These specific considerations show
up only at the stage where we display the explicit matrix elements
of the effective two-level hamiltonian. The following analysis is,
however, independent of the explicit form of the overall
perturbation $H_1$.

\subsection{\label{sec:WWpert} Mass matrix}

We have a system described by the Schr\"{o}dinger wave function
$|\Psi(t)>$ whose time evolution is given by $(\hbar = 1)$
\begin{equation}\label{Schrodinger}
i \frac{d}{dt}|\Psi(t)> = (H_0 + H_1)|\Psi(t)>.
\end{equation}
We write the full Hamiltonian $H = H_0 + H_1$, here $|L>$ and $|R>$
are the two degenerate discrete eigenstates of $H_0$, that is, a
pair of mirror-image enantiomers, and the perturbation $H_1$ induces
transitions from these states to other (possibly unbound)
eigenstates $|k>$ of $H_0$ and possibly also between $|L>$ and
$|R>$. The problem to solve is the time evolution of a state
initially prepared as a superposition of the two degenerate chiral
states. In terms of the interaction representation, the state-vector
\begin{equation}\label{psi}
|\psi(t)> = e^{iH_0 t}|\Psi(t)>
\end{equation}
satisfies the equation
\begin{equation}\label{interactrep}
i \frac{d}{dt}|\psi(t)> = e^{iH_0 t}H_1e^{-iH_0 t}|\psi(t)> =
H_1'|\psi(t)>,
\end{equation}
so that the time dependence of $|\psi(t)>$ arises solely from the
perturbation term $H_1$. We expand the interaction representation
wave function in terms of the complete set of eigenstates:
\begin{equation}\label{psiexpand}
|\psi(t)> = a(t)|L> + b(t)|R> + \sum_k c_k(t)|k>,
\end{equation}
subject to the initial conditions $a(0)=a_0, b(0) = b_0$, and
$c_k(0) = 0$. Then Eq. (\ref{interactrep}) leads to the following
set of coupled equations for the probability amplitudes:
\begin{eqnarray}\label{amplitudea}
i\frac{d a(t)}{dt} &=& <L|H_1|L> a(t) + <L|H_1|R> b(t) +
\sum_k <L|H_1'|k>c_k(t),\\
\label{amplitudebar-a} i\frac{d b(t)}{dt} &=& <R|H_1|L> a(t) + <
R|H_1|R> b(t) +
\sum_k <R|H_1'|k>c_k(t),\\
\label{amplitudec} i\frac{d c_k(t)}{dt} &=& <k|H_1'|L>a(t) +
<k|H_1'|R> b(t) + \sum_j <k|H_1'|j>c_j(t).
\end{eqnarray}
We may omit the prime on $H_1$ for the matrix elements taken within
the $|L>,|R>$ subspace in Eqs. (\ref{amplitudea}) and
(\ref{amplitudebar-a}) , since these are assumed to be degenerate in
mass.
Applying the Weisskopf-Wigner \cite{WW} approximation procedure to
this multi-level system yields the \textit{effective} two-level
quantum-mechanical description (see Eq. \ref{WWapprox})
\begin{eqnarray}\label{MGamma}
i\frac{d}{dt}\bm{\Phi} &=& \Big( \mathbf{M} - i \bm{\Gamma}
\Big)\bm{\Phi}.
\end{eqnarray}
The hermitian mass and decay matrices $\mathbf{M}=
\mathbf{M}^{\dagger}$ and $\bm{\Gamma} = \mathbf{\Gamma}^{\dagger}$
have the explicit matrix elements at order $O(H_1^2)$
\begin{equation}\label{Mmatrix}
M_{\alpha \beta} = m\delta_{\alpha \beta} +<\alpha|H_1|\beta>
-PP\sum_k \frac{<\alpha|H_1|k><k|H_1|\beta>}{E_k - m},
\end{equation}
and
\begin{equation}\label{Gmatrix}
\Gamma_{\alpha \beta} = 2\pi \sum_k
<\alpha|H_1|k><k|H_1|\beta>\delta(E_k - m),
\end{equation}
where the indices $\alpha, \beta$ stand for the states $L$ or $R$.

The only assumptions that go into obtaining the result in Eqs.
(\ref{MGamma})--(\ref{Gmatrix}) are that the dynamics is determined
by the time-dependent Schr\"{o}dinger equation, the higher order
terms $O(H_1^3)$ are neglected, and that the Hamiltonian $H$ is
hermitian (see the Appendix for details). This result allows for the
possibility of energy-conserving decay channels through the decay
matrix $\bm{\Gamma}$. Assuming that the chiral enantiomers are not
\textit{unstable}, there will be no decay, that is, no fragmentation
nor dissociation of the enantiomers into other molecular species.
Barring this possibility, there will be no contribution from Eq.
(\ref{Gmatrix}) because $E_k > m$. One might nevertheless be tempted
to think that $\mathbf{\Gamma}$ could automatically account for
effects of elastic collisions. Indeed, the full operator structure
$\mathbf{M} - i \bm{\Gamma}$ of the right hand side of
Eq.(\ref{MGamma}) implies complex energies for the spectrum, and is
reminiscent of the terms that are added \textit{phenomenologically}
to the molecular hamiltonian as a way of simulating approximately
the effects of collisions and radiative effects \cite{Quacka}.
However, its inclusion would imply exponential decay in the
probabilities $P_L(t), P_R(t)$ themselves, as well of course in the
optical activity $\sim (P_L-P_R)$,  so that e.g., $P_L(t),P_R(t)
\rightarrow 0$ as time increases, whereas collisional effects yield
instead $P_L(t),P_R(t) \rightarrow \frac{1}{2}$ \cite{Cattani}.
Henceforth, we set $\bm{\Gamma} = 0$ in the remainder of this paper.

\subsection{\label{sec:sub1} Eigenvalues and eigenvectors: \textit{CPT}
and T-invariance}

The underlying assumed invariance affects the form of the
eigenvalues and eigenvectors of Eq.(\ref{MGamma}) and it is
therefore of interest to consider independently the implications of
first \textit{CPT} and then T invariance.  The former applies only
for the case of the so-called $CP$-enantiomers, which require the
existence of mirror image molecules composed of
\textit{antiparticles} \cite{Barron1986}, that is, the CP partner of
$|L>$ is $|\bar R>$, the anti-right handed molecule, whereas the CP
partner of $|R>$ is $|\bar L>$, the anti-left handed molecule.

Thus the solution of the eigenvalue problem
\begin{equation}\label{eigenproblem}
\mathbf{M} \bm{\Psi}_{\pm} = \lambda_{\pm}\bm{\Psi}_{\pm},
\end{equation}
assuming \textit{CPT} invariance, so that $M_{11}=M_{22}$ (and
$\Gamma_{11} = \Gamma_{22}$) \cite{TDLee}, is given by
\cite{LeeOehmeYang}
\begin{equation}\label{eigenvectors}
|\bm{\Psi}_{\pm}> = \left(
                    \begin{array}{c}
                      p \\
                      \pm p^* \\
                    \end{array}
                  \right) \frac{1}{\sqrt{2|p|^2}},
\end{equation}
where
\begin{equation}
\lambda_{\pm} = M_{11} \pm |M_{12}|,
\end{equation}
and $p$ is the complex number
\begin{equation}
p^2  = M_{12}.
\end{equation}
From Eq. (\ref{eigenvectors}), and in terms of the CP-enantiomers we
have
\begin{eqnarray}
|L> &=& \frac{e^{i\alpha}}{\sqrt{2}}\Big( |\bm{\Psi}_{+}> +
|\bm{\Psi}_{-}> \Big),\\
|\bar R> &=& \frac{e^{i\alpha}}{\sqrt{2}}\Big( |\bm{\Psi}_{+}> -
|\bm{\Psi}_{-}> \Big),\\
\end{eqnarray}
where $e^{-i\alpha} = p/|p|$.

If on the other hand we assume only T-invariance, then $M_{12}^* =
M_{12}$ \cite{TDLee}, and in this case the eigenvalues and
eigenvectors of Eq.(\ref{eigenproblem}) are given by
\cite{Barron1986}
\begin{equation}\label{eigenvaluesT}
\lambda_{\pm} = \frac{1}{2}(M_{11}+M_{22}) \pm
\frac{1}{2}[(M_{11}-M_{22})^2 + 4M_{12}^2]^{1/2},
\end{equation}
\begin{eqnarray}\label{eigenvectors2a}
|\bm{\Psi}_{+}> &=& \cos \phi |L> + \sin \phi |R>,\\
\label{eigenvectors2b} |\bm{\Psi}_{-}> &=& -\sin \phi |L> + \cos
\phi |R>,
\end{eqnarray}
where $\cot 2\phi = \frac{(M_{11}-M_{22})}{2|M_{12}|}$.

\subsection{\label{sec:optical} Corrections to the optical
activity}

Now we can determine how the inclusion of multiple states affects
the optical activity with respect to the simplest HS model. We
prepare the state to be initially $|L>$ and assume first only T
invariance. Using Eqs. (\ref{eigenvaluesT})--(\ref{eigenvectors2b}),
we find the wavefunction at any time is given by
\begin{equation}
|\Psi(t)> = (\cos^2 \phi e^{-i\lambda_{+}t} + \sin^2 \phi
e^{-i\lambda_{-}t})|L> + \sin \phi \cos \phi (e^{-i\lambda_{+}t} -
e^{-i\lambda_{-}t}) |R>.
\end{equation}
Following a sequence of steps similar to those in Sec \ref{sec:HS},
we can evaluate the probabilities $P_L,P_R$ and the optical activity
in the presence of multi-state transitions. The probabilities to be
in state $|L>$ or $|R>$ at any time $t \geq 0$ are given by
\begin{eqnarray}
P_L(t) &=& \cos^2 (\Delta t) + \frac{(M_{11}-M_{22})^2}{4 \Delta^2}
\sin^2 (\Delta t), \\
P_R(t) &=& \Big( \frac{M_{12}^2}{\Delta^2} \Big) \sin^2 (\Delta t).
\end{eqnarray}

We find that
\begin{equation}\label{OAcorrections1}
\Theta(t) = \Theta_{max}\frac{\frac{1}{4}(M_{11}-M_{22})^2 +
M_{12}^2 \cos(2 \Delta t/\hbar)}{\Delta^2},
\end{equation}
where the explicit matrix elements $M_{\alpha \beta}$ are given by
Eq.(\ref{Mmatrix}) and $\Delta = \frac{1}{2}[(M_{11}-M_{22})^2 +
4M_{12}^2]^{1/2}$.

When \textit{CPT} invariance holds $M_{11}=M_{22}$, and then the
optical activity is given by
\begin{equation}\label{OAcorrections2}
\Theta(t) = \Theta_{max} \cos(2 \Delta t/\hbar),
\end{equation}
where now, $\Delta = |M_{12}|$. If we denote the oscillation periods
$\tau_{\textit{CPT}}$ and $\tau_{T}$ when \textit{CPT} or T
invariance is imposed, then we have the general result that
\begin{equation}
\tau_{\textit{CPT}} > \tau_{T},
\end{equation}
so that the oscillation period in the case of the hypothetical
CP-enantiomers is \textit{longer} then that for the P-enantiomers,
for a given interaction hamiltonian $H_1$.

The time-average optical activity for the case of P-enantiomers is
\begin{equation}
\left\langle \frac{\theta(t)}{\theta_{max}} \right\rangle_t =
\frac{\frac{1}{4}(M_{11} - M_{22})^2}{\frac{1}{4}(M_{11} - M_{22})^2
+ M_{12}^2}.
\end{equation}
If \textit{CPT} invariance is assumed, then the time-average of
Eq.(\ref{OAcorrections2}) is zero:
\begin{equation}
\left\langle \frac{\theta(t)}{\theta_{max}} \right\rangle_t = 0.
\end{equation}
As mentioned earlier, we write $H_1 = \delta \sigma_x + \epsilon
\sigma_z + H_2$, where the first two terms act only within the two
dimensional subspace of the enantiomers $|L>,|R>$ and $H_2$ induces
transitions from these to the other levels $|k>$, thus we evaluate
\begin{equation}\label{Mmatrix2}
M_{\alpha \beta} = m\delta_{\alpha \beta} +<\alpha|\delta \sigma_x +
\epsilon \sigma_z |\beta> -PP\sum_k
\frac{<\alpha|H_2|k><k|H_2|\beta>}{E_k - m}.
\end{equation}
If we shut off the perturbation $H_2$ then from Eq. (\ref{Mmatrix})
it is easy to check that Eq.(\ref{OAcorrections1}) reduces to the
optical activity of the isolated two-state system, Eq.(\ref{HSOA}).
By the same token, Eq.(\ref{OAcorrections2}) reduces to the
Eq.(\ref{HSOA1}) in this same limit.

Note that since $\mathbf{M}$ is Hermitian, $P_L(t) + P_R(t) = 1$
continues to hold, even allowing for transitions to the other states
$|k>$. If $\bm{\Gamma}$ were not vanishing, then these probabilities
would decay exponentially in time.

\section{\label{sec:kaon} Formal analogies to the \textit{K}-meson system}

In the context of symmetry breaking in physics, Wigner pointed out
some time ago a strictly formal analogy between \textit{K}-mesons
and chiral molecules \cite{Wigner1965}. The neutral $K^0$ meson and
its antiparticle $\bar{K}^0$ are related by the combined operations
of charge conjugation and parity (CP): $|\bar{K}^0> = CP |K^0>$
\cite{TDLee}. From this, one defines superpositions \cite{MGM}
$|K_1> = \frac{1}{\sqrt{2}}(|K^0> + |\bar{K}^0>)$ and $|K_2> =
\frac{1}{\sqrt{2}}(|K^0> - |\bar{K}^0>)$ that are eigenstates of CP:
$CP|K_1> = |K_1>$ and $CP|K_2> = -|K_2>$. The chiral molecules are
interrelated by the parity operation: $|L> = P|R>, |R> = P|L>$, and
the eigenstates of definite parity are the mixtures $|+> =
\frac{1}{\sqrt{2}}(|L> + |R>)$ and $|+> = \frac{1}{\sqrt{2}}(|L> -
|R>)$, as $P|+> = |+>$ and $P|-> = -|->$. These algebraic
relationships led Wigner to propose a formal analogy between neutral
kaons and enantiomers, namely, the state-vector associations
($\leftrightarrow$)
\begin{eqnarray}
|L> &\leftrightarrow& |K^0> ,\\
|R> &\leftrightarrow& |\bar{K}^0>,\\
|+> &\leftrightarrow& |K_1>,\\
|-> &\leftrightarrow& |K_2>.
\end{eqnarray}
This analogy can be made more encompassing, by extending these
relationships to the dynamic level. The kaons are eigenstates of the
strong $H_{st}$ and electromagnetic $H_\gamma$ interactions:
$(H_{st} + H_{\gamma})|K^0> = m_K |K^0>$ and $(H_{st} +
H_{\gamma})|\bar{K}^0> = m_K |\bar{K}^0>$ and are degenerate in
mass. The weak interaction $H_{weak}$ connects $K^0$ and $\bar{K}^0$
with other continuum states which causes the various decay modes and
removes their degeneracy. The Schr\"{o}dinger equation
Eq.(\ref{Schrodinger}), describes the time evolution of a neutral
kaon system, within the two-level approach in
Eqs.(\ref{MGamma})--(\ref{Gmatrix}), For this, one makes the
specific identifications
\begin{eqnarray}
H_0 &=& H_{st} + H_{\gamma}, \\
H_1 &=& H_{weak}, \\
\end{eqnarray}
where now the indices $\alpha, \beta$ stand for the states $K^0$ or
$\bar {K}^0$ and $m = m_K = m_{\bar K}$
\cite{LeeOehmeYang,TDLee,Kabir,Sachs}. Due to kaon decays, the decay
matrix $\bf{\Gamma}$ is nonzero. The eigenvalue problem for $(
\mathbf{M} - i \bm{\Gamma}) $ has been worked out in full detail
\cite{LeeOehmeYang}. The eigenvalues have real and imaginary parts,
and these can be expressed as $m_{1,2} - \frac{i}{2}\gamma_{1,2}$
\cite{TDLee}. So if we prepare a state which is initially pure
$K^0$, at any later time the probability to find a $\bar{K}^0$ is
\cite{Kabir} ($\hbar = 1$)
\begin{eqnarray}
P(\bar{K}^0,t) &=& \frac{1}{4}[e^{-\gamma_1 t} + e^{-\gamma_2 t}
-2e^{-(\gamma_1+ \gamma_2) t}\cos(m_2-m_1)t],\\
&=& \cos^2 \Big(\frac{(m_2-m_1)t}{2}\Big),\qquad (\gamma_{1,2}
\rightarrow 0),
\end{eqnarray}
which describe decaying meson oscillations. In the limit of zero
decay (second line) these would become pure oscillations and thus
formally similar to the chiral oscillations derived above. These
considerations raise Wigner's static analogy relating state vectors,
to a \textit{dynamic} one, between effective two-level
Schr\"{o}dinger equations, oscillations in the transition
probability between $L$ and $R$, Eq. (\ref{LtoRtrans}), and the
strangeness oscillations of the neutral kaon system.

We must point out however that the consideration given to a
two-state model involving CP-enantiomers in Secs \ref{sec:sub1} and
\ref{sec:optical} is strictly a mathematical illustration only and
can never be realized experimentally. This is because, such
molecule-antimolecule transformations would require a huge violation
of baryon number conservation \cite{Barron1994}. This problem does
not arise in the K-meson system because mesons have baryon number
zero (mesons are not baryons!)

\section{\label{sec:disc} Discussion}

A number of criticisms of the simplest two-state HS model have been
marshalled in the past: namely that it could only apply at
exceedingly low temperatures, that it neglects the radiation field,
and that it does not account for collisions. It has been suggested
that perhaps the most serious problem arises from the multi-state
nature of real molecules \cite{Quacka}. It is the latter objection
which motivates the work presented here, providing us the incentive
to consider the multi-state nature in a fairly general way. One
aspect (and only one) of this complex problem is the influence of
transitions to a tower of excited (electronic-vibrational) states of
enantiomers induced by an appropriate external field or
perturbation. In this situation, results from Weisskopf-Wigner
perturbation theory demonstrate that we can continue to employ an
\textit{effective} two-state description, where the influence of the
tower of multiple states is accounted for by the matrix elements of
the effective hamiltonian or mass matrix acting in the subspace of
the two enantiomers. The effects that multiple states have on the
racemization and optical activity can then be worked out in terms of
the explicit matrix elements of the specific interaction responsible
for these transitions. The importance of assuming \textit{CPT} or
T-invariance is underscored here. These results hold generally. The
main result is that a two-state approach remains valid, because the
inclusion of the multiple states can be included in an effective
Hamiltonian description. An approach such as this may prove useful
for interpreting proposed spectroscopic measurements of molecular
parity violation, such as represented in Fig. 1 of reference
\cite{Quack2008}, involving transitions to excited levels.

In comparison to the general results obtained here, much of the
theoretical work has focused on the explicit calculations of the
parity violating energy difference in chiral molecules; see
\cite{Quack2008} for a recent review. These involve ab initio
computations of the PV interactions employing techniques such as
non-relativistic and relativistic (Dirac)-Hartree-Fock and
multi-configurational self-consistent (MCSCF) levels, as well as
density functional theory (DFT) \cite{Bakasov,Lazzeretti,Laerdahl,
Thyssen,Bergerc,Bergerd,Schwerdtfeger,Hennum}. There the primary
objective is the (numerical) evaluation of effective
parity-violating Hamiltonians which requires using many-body
quantum-mechanical wave functions, to account for the multiple
nuclei and electrons \cite{Quack2008} involved. In these
investigations, the dominant contribution to the parity-violating
energy difference between enantiomers $E_{pv}$ is calculated from
matrix elements connecting the ground-state singlet with excited
triplet states. Thus, it should be possible to employ an effective
two-level description in these more complex theoretical approaches
as well.

Regarding collisional effects, a two-level approach has been used to
describe how racemization depends on the interaction of the
enantiomers with the environment \cite{Cattani}.  This might suggest
that both the multi-state nature of real molecules and collisional
effects with the surrounding medium might be able to be combined in
an overall effective two-level description.

Finally, we have also further developed the formal dynamical
analogies between the system of enantiomers and the kaon system. The
unifying framework is provided by the WW perturbation theory. We
note that a formal comparison between chiral molecules and neutrinos
was recently invoked to derive properties of the oscillations
between isolated enantiomers in a two-level HS-type approximation
\cite{Bargue}.

\begin{acknowledgments}
The research of DH is supported in part by the Grant
AYA2009-13920-C02-01 from the Ministerio de Ciencia e Innovaci\'{o}n
(Spain) and forms part of the COST Action CM0703 ``Systems
Chemistry". C.B. acknowledges financial support from the Instituto
Nacional de T\'{e}cnica Aeroespacial (INTA). We thank Isabel Gonzalo
for providing C.B. with bibliography on racemization and parity
violation in molecules.
\end{acknowledgments}

\appendix
\section{\label{sec:WW} Reduction from multi-level to a two-state system}

The main steps of the Weisskopf-Wigner (WW) time dependent
perturbation theory \cite{WW,Heitler,LeeOehmeYang} and Appendix A of
\cite{Kabir}, are reviewed and adapted here to a doublet of
mass-degenerate enantiomers. We emphasize that the WW method is
general and provides a way to reduce an \textit{a-priori} multiple
level quantum system to an effective two-level system, independent
of the actual form of the specific hermitian Hamiltonians involved.

To derive Eq.(\ref{MGamma}) from Eqs.
(\ref{amplitudea})--(\ref{amplitudec}), introduce the two-component
column vectors
\begin{equation}
\bm{\phi}(t) = \left(\begin{array}{c}
                 a(t) \\
                 b(t)
               \end{array}\right),
               \, \mathbf{C}_k = \left(\begin{array}{c}
                 <k|H_1|L> \\
                 <k|H_1|R>
               \end{array}\right),
\end{equation}
then the first WW approximation consists in truncating the solution
of the Eqs. (\ref{amplitudea})--(\ref{amplitudec}) to second order
in $H_1$. This implies they can be written as follows, by using
$<k|H_1'|L> = e^{i w_k t} <k|H_1|L>, <k|H_1'|R> = e^{i w_k t}
<k|H_1|R>$:
\begin{eqnarray}\label{WWapprox1a}
i\frac{d \bm{\phi}(t)}{dt} &=& \mathbf{h}\bm{\phi}(t) + \sum_k
\mathbf{C}^*_k e^{-iw_k t}c_k(t),\\ \label{WWapprox1b} i\frac{d
c_k(t)}{dt} &=& \mathbf{C}^T_k \bm{\phi}(t)e^{iw_k t},
\end{eqnarray}
where $w_k = E_k - m$, $(m_L = m_R \equiv m)$ and $\mathbf{h}$ is
the submatrix of $H_1$ in the two-state subspace:
\begin{equation}
\mathbf{h} = \left(
               \begin{array}{cc}
                 <L|H_1|L> &   <L|H_1|R> \\
                 <R|H_1|L> & <R|H_1|R> \\
               \end{array}
             \right).
\end{equation}

Solve Eq.(\ref{WWapprox1b}) for $c_k$ and substitute these solutions
back into Eq.(\ref{WWapprox1a}). The resultant equation for
$\bm{\phi}(t)$ can be solved in closed form via a Laplace transform
\cite{Kabir} and yields
\begin{equation}\label{solution}
\bm{\phi}(t) = \frac{1}{2\pi i}\int_{-\infty}^{\infty} dy
\frac{e^{(iy + \epsilon)t}}{y-i\epsilon + \mathbf{W}(iy+\epsilon)
}\bm{\phi}_0,
\end{equation}
where $\bm{\phi}_0 = \bm{\phi}(0)$ is the initial condition, and
\begin{equation}\label{W}
\mathbf{W}(s) = \mathbf{h} - \sum_k \frac{\mathbf{D}_k}{w_k - is},
\qquad \mathbf{D}_k = \mathbf{C}_k^*\mathbf{C}_k^T.
\end{equation}
Up to this point, the solution Eq.(\ref{solution}) is exact to
$O(H_1^2)$. If we regard the perturbation $H_1$ as small, then the
second order contribution to the matrix $\mathbf{W}$ should receive
its main contribution to the integral from the neighborhood of
$y=0$. The second WW approximation consists in replacing
$\mathbf{W}$ by its value at $y=0$, which leads to the integral
\begin{equation}\label{integral}
\bm{\phi}(t) = \frac{1}{2\pi i} \int_{-\infty}^{\infty} dy \,
e^{iyt} \Big(y + \mathbf{h} -PP\sum_k \frac{\mathbf{D}_k}{w_k} -i\pi
\sum_k \delta(w_k) \mathbf{D}_k \Big)^{-1} \bm{\phi}_0,
\end{equation}
and which follows from the identity ($PP$ denotes the Cauchy
principal part) \cite{Heitler}
\begin{equation}
\lim_{\sigma \rightarrow 0} \frac{1}{x \pm i \sigma} = \frac{PP}{x}
\mp i\pi \delta(x).
\end{equation}
Evaluating the integral Eq. (\ref{integral}) yields the general
solution
\begin{equation}
\bm{\phi}(t) = e^{-i\mathbf{W}_0 t}\bm{\phi}_0,
\end{equation}
where
\begin{equation}
\mathbf{W}_0 = \mathbf{h} -PP\sum_k \frac{\mathbf{D}_k}{w_k} -i\pi
\sum_k \delta(w_k) \mathbf{D}_k.
\end{equation}

The time dependence in the interaction representation of the
two-level wavefunction in the WW approximation is given by
\begin{equation}
i\frac{d}{dt}\bm{\phi}(t) = \mathbf{W}_0 \bm{\phi}(t).
\end{equation}
Returning now to the Schr\"{o}dinger representation $\bm{\Phi} =
e^{-i H_0 t} \bm{\phi}$,
\begin{eqnarray}
i\frac{d}{dt}\bm{\Phi} &=& (H_0 + e^{-iH_0 t} \mathbf{W}_0 e^{iH_0
t}) \bm{\Phi},\\
&=& (H_0 + \mathbf{W}_0 ) \bm{\Phi},\\ \label{WWapprox} &=& \Big(
\mathbf{M} - i \bm{\Gamma} \Big)\bm{\Phi}.
\end{eqnarray}
Here
\begin{equation}\label{M}
\mathbf{M} = m \mathbf{1} + \mathbf{h} -PP\sum_k
\frac{\mathbf{D}_k}{w_k},
\end{equation}
and
\begin{equation}\label{Gamma}
\bm{\Gamma} = 2\pi \sum_k \delta(w_k) \mathbf{D}_k ,
\end{equation}
are known as the mass and decay matrices, respectively.  Since the
two level subsystem is degenerate, $H_0 = m\mathbf{1}$ and so
$[e^{-iH_0 t}, \mathbf{W}_0] = 0$.

\end{document}